# AI Regulation in Europe: From the AI Act to Future Regulatory Challenges

Philipp Hacker[*]

**Abstract:**

This chapter provides a comprehensive discussion on AI regulation in the European Union, contrasting it with the United Kingdom's more sectoral and self-regulatory approach. It argues for a hybrid regulatory strategy that combines elements from both philosophies, emphasizing the need for agility and safe harbours to ease compliance. The paper examines the EU's AI Act as a pioneering legislative effort to address the multifaceted challenges posed by AI, asserting that, while the Act is a step in the right direction, it has shortcomings that could hinder the advancement of AI technologies. The paper also anticipates upcoming regulatory challenges, such as the management of toxic content, environmental concerns, and hybrid threats. It advocates for immediate action to create protocols for regulated access to high-performance, potentially open-source AI systems. Although the EU's AI Act is a significant legislative milestone, it needs additional refinement and global collaboration for the effective governance of rapidly evolving AI technologies.

Keywords: AI Act; artificial intelligence; foundation models; product liability; sustainability; toxicity; threats

**Contents:**

---

[*] Prof. Dr. Philipp Hacker, LL.M. (Yale), holds the Chair for Law and Ethics of the Digital Society at European New School of Digital Studies at the European University Viadrina.

## I. Introduction

Artificial intelligence (AI) and its most important current instantiation, machine learning, have made enormous progress in recent years. Since OpenAI introduced ChatGPT in December 2022 and the enhanced version GPT-4 shortly afterwards, AI has become part of many people's everyday lives. Even before that, AI has powered many important economic and administrative applications, from face recognition to cancer detection, and from spam control to handwriting deciphering and climate change mitigation tools. With the advent of generative AI systems, such as ChatGPT, Bard, or Stable Diffusion, AI is being deployed to every corner of our society as an unprecedented pace.

This trend harbours significant potential, but also raises concerns. Much has been written about AI risks, from opacity to discrimination, data protection violations and unforeseeability. More recently, the environmental costs of training large AI systems–GHG emissions, water consumption, and toxic materials–have come under increased scrutiny in the machine learning community. Moreover, a controversial discourse around existential risks triggered by an abuse of AI by malicious actors or autonomously acting wrote AI has arisen.

Given these developments, it is not surprising that legislators and regulators around the world are trying to square the regulatory circle by containing AI's risks without stifling innovation and beneficial use. In this context, the European Commission first proposed a comprehensive legal framework to regulate the use of AI in its Member States in April 2021: the *Artificial Intelligence Act* (AI Act). The AI Act aims to increase trust in AI and ensure that this technology is used in a way that respects the fundamental rights and safety of EU citizens.

However, the EU is no longer alone in its regulatory agenda. China has already passed an AI law that is structured in a sectoral way and even provides regulations for generative AI.[1] Canada is preparing a corresponding law. The US Senate has held extensive hearings, even if the House of Representatives is dysfunctional at this moment and the likelihood of regulation at the federal level is very low. Brazil, on the other hand, is taking its cue from the AI Act. Internationally, too, with the G7 Hiroshima AI Process and initiatives at the United Nations level, various projects are underway, ranging from voluntary commitments to a possible international AI convention.

The present chapter will explore, against this background, how the Europe positions itself in this dynamic environment.[2] The clear focus will be on the EU, but developments in the UK will be surveyed as well. Importantly, the AI Act is only one mosaic stone among many. The revision of the Product Liability Directive is potentially even more important for legal practice and innovation.[3] A new proposal, bound to be enacted jointly with the AI Act, will extend the European product liability framework to AI and software; and include extensive disclosure obligations as well as a reversal of evidence with regard to product defects and causality, which would apply in particular to machine learning applications. Even though this article focuses on the AI Act, the downstream effects triggering civil liability in the event of a breach of the AI Act must always be considered. Moreover, the Digital Services Act has recently come into effect, technically regulating platforms and other intermediaries,[4] but potentially also tackling certain types of hate speech and fake news generated by AI systems.[5] The Digital Markets Act, geared towards strengthening competition and the platform economy,[6] also applies to AI used by large online platforms, for example, when ranking products in response to search queries.[7] Finally, technology-neutral laws, such as the prohibition of discrimination[8] and the GDPR,[9] continue to apply to AI systems.

---

[1] See Matt O'Shaughnessy and Matt Sheehan, Lessons from the World's Two Experiments in AI Governance, CEIP (Feb. 14, 2023), https://carnegieendowment.org/2023/02/14/lessons-from-world-s-two-experiments-in-ai-governance-pub-89035.
[2] See also, for a German treatment, Philipp Hacker and Amelie Berz, Der AI Act der Europäischen Union – Überblick, Kritik und Ausblick, ZRP (2023), (forthcoming).
[3] See, e.g., Philipp Hacker, 'The European AI Liability Directives - Critique of a Half-Hearted Approach and Lessons for the Future' (2023) 51 Computer Law & Security Review 51 (2023), Article 105871; Christiane Wendehorst, *AI liability in Europe: anticipating the EU AI Liability Directive* (Ada Lovelace Institute 2022).

While the EU has responded with a whole range of regulatory initiatives to the challenges raised by AI, the UK is following a distinctly different approach. Section II of the chapter will therefore briefly contrast the EU and the UK modes of regulation, before Section III presents the general architecture and contents of the EU AI Act. Section IV situates the act within the broader international and economic environment. Section V offers a critique of this landmark act and make policy suggestions. Finally, Section VI identifies three core challenges that future AI regulation, in the EU and beyond, should focus in to rein in AI's externalities: toxicity; environmental sustainability; and hybrid threats, facilitated particularly by powerful open-source models. Section VII. concludes.

## II.     Modes of regulation: the EU versus the UK?

The European Union and the United Kingdom have adopted distinct approaches towards the regulation of AI, reflecting their respective governance philosophies and priorities. The EU largely opts for a "command-and-control" regulatory style, paired with a risk-based approach, notably in its AI Act. In it, the EU combines broad regulatory obligations with partially external conformity assessment with self-certification for various classes of AI systems.[10] This approach is bolstered by an ambitious and far-reaching update of the product liability framework.[11]

---

[4] See, e.g., Folkert Wilman, 'The Digital Services Act (DSA)-An Overview' (2022) Available at SSRN 4304586; Caroline Cauffman and Catalina Goanta, 'A new order: The Digital Services Act and consumer protection' (2021) 12 European Journal of Risk Regulation 758.

[5] See, e.g., Natali Helberger and Nicholas Diakopoulos, 'ChatGPT and the AI Act' (2023) 12 Internet Policy Review; Philipp Hacker, Andreas Engel and Marco Mauer, 'Regulating ChatGPT and other Large Generative AI Models' (2023) ACM Conference on Fairness, Accountability, and Transparency (FAccT '23) 1112, 1117.

[6] See, e.g., Martin Eifert and others, 'Taming the giants: The DMA/DSA package' (2021) 58 Common Market Law Review 987.

[7] See, e.g., Philipp Hacker, Johann Cordes and Janina Rochon, 'Regulating Gatekeeper AI and Data: Transparency, Access, and Fairness under the DMA, the GDPR, and beyond' (2022) Working Paper, https://arxivorg/abs/221204997.

[8] See, e.g., Jeremias Adams-Prassl, Reuben Binns and Aislinn Kelly-Lyth, 'Directly Discriminatory Algorithms' (2022) The Modern Law Review 144; Sandra Wachter, 'The Theory of Artificial Immutability: Protecting Algorithmic Groups Under Anti-Discrimination Law' (2022) arXiv preprint arXiv:220501166.

[9] See, e.g., Michael Veale, Reuben Binns and Lilian Edwards, 'Algorithms that remember: model inversion attacks and data protection law' (2018) 376 Philosophical Transactions of the Royal Society A: Mathematical, Physical and Engineering Sciences 20180083; Michael Butterworth, 'The ICO and artificial intelligence: The role of fairness in the GDPR framework' (2018) 34 Computer Law & Security Review 257; Philipp Hacker, 'A legal framework for AI training data—from first principles to the Artificial Intelligence Act' (2021) 13 Law, Innovation and Technology 257.

[10] For a critique, see, e.g., Michael Veale and Frederik Zuiderveen Borgesius, 'Demystifying the Draft EU Artificial Intelligence Act—Analysing the good, the bad, and the unclear elements of the proposed approach' (2021) 22 Computer Law Review International 97.

[11] European Commission, Proposal for a Directive of the European Parliament and of the Council on Liability for Defective Products (2022) [PLD Proposal]; European Commission, Proposal for a Directive of the European

Specifically, the new regime bound to take effect jointly with the AI Act stipulates that for any AI applications, providers must fulfil extensive evidence disclosure requirements. This even extends to reversals of the burden of proof in liability cases, where the onus is shifted onto the AI provider to demonstrate the absence of a product defect in cases of machine learning.

In contrast, the UK has leaned more towards a self-regulatory model, particularly under its current government, which emphasizes AI safety and existential risk.[12] Building largely on existing sectoral regulation,[13] the regulatory environment is less prescriptive, giving technology companies greater leeway to innovate while aiming to still adhere to general principles of safety. This manifests in various guidelines and voluntary standards that firms are encouraged to follow but are not legally mandated to do so. The UK's focus on existential risks implies a long-term view that considers not just the immediate impact of AI but its potential future developments, including the theoretical advent of Artificial General Intelligence (AGI). Put positively, the UK government aims to cultivate a regulatory ecosystem that is both flexible and anticipatory; put negatively, it neglects current AI risks, such as discrimination, opacity, toxicity, and unforeseeable output.[14]

These contrasting regulatory frameworks arguably reflect broader political divergences between the EU and the UK, particularly in the realms of market intervention, consumer protection, and innovation policy. Very broadly sketched, the EU's approach is rooted in its broader social market economy,[15] where regulation often serves as a tool for harmonizing market conditions while aiming to ensure high levels of consumer protection and fundamental rights. The UK, on the other hand, generally adheres to a more liberal market philosophy that prioritizes innovation and economic competitiveness, albeit with certain safety considerations.[16] Both models offer their own sets of advantages and challenges, and it remains to be seen how each will adapt in the face of rapid advancements in AI technology.

### III. Architecture and main content of the AI Act

The most imposing and ambitious piece of AI regulation in Europe is undeniably the European Union's AI Act. Originally proposed by the European Commission in April 2021,[17] the Act aims to create a comprehensive and harmonized framework for the development, deployment, and oversight of artificial intelligence across the EU. In December 2022, the Council of the EU

---

Parliament and of the Council on adapting non-contractual civil liability rules to artificial intelligence (2022) [AILD Proposal]; see also n 3.
[12] See https://www.gov.uk/government/news/uk-government-sets-out-ai-safety-summit-ambitions.
[13] Huw Roberts and others, 'Artificial intelligence regulation in the United Kingdom: a path to good governance and global leadership?' (2023) 12 Internet Policy Review, DOI: 10.14763/2023.2.1709.
[14] See Sandra Wachter and Brent Mittelstadt, 'No need to wait for the future: The danger of AI is already here' Oxford Internet Institute Blog (May 15, 2023) <https://www.oii.ox.ac.uk/news-events/news/no-need-to-wait-for-the-future-the-danger-of-ai-is-already-here/>.
[15] Christian Joerges and Florian Rödl, 'Social market economy as Europe's Social Model' in Lars Magnusson and Bo Stråth (eds), *A European social citizenship* (2004), 125; Dragana Damjanovic, 'The EU Market Rules as Social Market Rules: Why the EU can be a social market economy' (2013) 50 Common Market Law Review 1685.
[16] For more nuance, see, e.g., Iain McLean, 'The history of regulation in the United Kingdom: Three case studies in search of a theory' in David Levi-Faur Jacint Jordana (ed), *The Politics of Regulation* (2004), 45.
[17] European Commission, Proposal for a Regulation of the European Parliament and of the Council Laying Down Harmonised Rules on Artificial Intelligence (Artificial Intelligence Act), COM(2021) 206 final.

representing the Member States adopted its position, the so-called general approach.[18] With generative AI moving to the front line of the legislative awareness, the European Parliament deliberated until June 2023 to propose its amendments,[19] which contain dedicated sections on foundation models and generative AI (see below, V.).

While being a key piece of EU legislation, the AI Act unfolds an extraterritorial effect, just like the GDPR.[20] Irrespective of where exactly a provider is based or the AI has been trained, the AI Act applies if the AI or its result is used in the EU (Art. 2(1) AI Act). EU legislators, therefore, cast their net widely. However, as can be noticed currently with the DSA, the extraterritorial effect may lead to regional rollouts of technology: Meta and other companies are creating GDPR and DSA-compliant versions, facilitating opt outs from tracking and personalization, for EU customers only, while preserving their non-compliant versions in the rest of the world. A similar reaction can be expected with respect to the AI Act, so that the achievement of a "Brussels effect"[21] remains doubtful.[22]

Importantly, again like the GDPR,[23] the AI Act follows a risk-based approach.[24] It distinguishes between four different risk categories prohibited; high-risk; limited-risk; and unregulated. Art. 5 AI Act lists prohibited applications of AI, such as social scoring or (with highly controversial exceptions) biometric identification measures in public spaces. The heart of the planned AI regulation, however, are the rules for so-called high-risk AI systems (Art. 6 et seqq. AI Act).

For high-risk AI systems, providers must:

- Conduct risk assessments and maintain high-quality datasets
- Make available extensive documentation and keep records
- Maintain transparency by providing adequate information to users
- Ensure human oversight during the AI system's operation
- Implement robustness, performance, and cybersecurity measures

High-risk areas include, for example, administration and justice, but also face recognition, medicine, employment, credit scoring and certain insurance contracts (life and health).

---

[18] Council of the EU, Interinstitutional File: 2021/0106(COD), General approach of Nov. 25, 2022, Doc. No. 14954/22, https://data.consilium.europa.eu/doc/document/ST-14954-2022-INIT/en/pdf (= AI Act Council Version).
[19] Draft Compromise Amendments on the Draft Report, Proposal for a regulation of the European Parliament and of the Council, Brando Benifei & Ioan-Dragos ̦ Tudorache (May 9, 2023), https://www.europarl.europa.eu/meetdocs/2014_2019/plmrep/COMMITTEES/CJ40/DV/2023/05-11/ConsolidatedCA_IMCOLIBE_AI_ACT_EN.pdf (= AI Act EP Version).
[20] See, e.g., Christopher Kuner, 'Territorial Scope and Data Transfer Rules in the GDPR: Realising the EU's Ambition of Borderless Data Protection' (2021) University of Cambridge Faculty of Law Research Paper.
[21] Coined by Anu Bradford, 'The Brussels Effect' (2012) 107 Nw UL Rev 1.
[22] See also Anu Bradford, *Digital Empires: The Global Battle to Regulate Technology* (Oxford University Press 2023).
[23] Raphael Gellert, 'Understanding the notion of risk in the General Data Protection Regulation' (2018) 34 Computer Law & Security Review 279.
[24] See, e.g., Hacker, Engel and Mauer, 'Regulating ChatGPT and other Large Generative AI Models'.

Third, systems that are not used in these areas only have to comply with Article 52 AI Act: If they interact with humans, the use of AI must be made transparent. This also applies to AI-powered chatbots and deepfakes. For all other systems, fourth, no special requirements apply.

This architecture evidences a central characteristic of the AI Act as it was originally conceived: its risk-based and use-case-oriented framework. The higher the risk to health, safety or fundamental rights posed by a specific application, the stricter the obligations. Moreover, as the AI Act was strongly inspired by product safety legislation, conformity assessments play a central role in the approval process. In many areas, however, developers can resort to self-certification.[25] Non-compliance with the Act can result in hefty fines, ranging up to 6 or 7% of the annual global turnover of the AI provider or user, depending on the nature and severity of the infringement.

### IV. International and economic considerations

For a critical discussion of the AI Act, the international context and the possible economic impact must be considered as well. A study from 2022 shows that in the six preceding years, 73 % of the large foundation models were published by US companies and 15 % in China.[26] This leaves less than 10% for the EU concerning this crucial sector of frontier AI systems. The result is alarming: after gas and oil, a new dependency threatens to emerge in a highly dynamic and confrontational geostrategic environment with regard to a key technology of the 21st century.

It should also be noted that many of the companies successfully developing advanced AI systems in the EU are SMEs (e.g., Aleph Alpha, Mistral, Poolside, nyonic). Compliance costs, however, are largely independent of company size. Efforts needed for assessing and mitigating risks are a function of model complexity and existing compliance infrastructure. Hence, compliance costs are generally much more difficult to bear for less financially strong SMEs than for Google, Microsoft and other large technology companies. This effect has already been noted with respect to the GDPR,[27] and will probably be reproduced for the AI Act. The EU, therefore, must be mindful not to squeeze out the few internationally competitive AI companies it has.

It therefore seems important, first, to maintain a use-case-oriented focus and, second, to pair the AI Act with specific support for SMEs in the digital sector: Unbureaucratic financial assistance and possible subsidies for any insurance may help. This is particularly important against the backdrop of considerable liability risks in conjunction with revised product liability. Third, it is essential that guidelines for the application of the AI Act are issued in rapidly and that *safe harbours* are defined as quantitatively as possible in order to reduce legal uncertainty. This is eminently important, especially for SMEs in the early stages: such companies tend to be backed by venture capitalists, who are often risk-averse with regard to legal disputes and quickly withdraw funding and support as soon as cases brought against the SMEs in court–irrespective of whether the case ultimately has merit or not. It should, therefore, be a priority for lawmakers and regulators to ensure that SMEs may achieve effective compliance, by taking appropriate technical precautions and adhering to safe harbours, to prevent potentially ruinous lawsuits. The possibility to self-certify AI Act compliance only marginally mitigates this risk as cases can, of course, be brought against companies regardless of whether they have self-certified as compliant or not. In conjunction, the AI Act and the liability package raise incentives for rational regulatory arbitrage,[28] i.e., the relocation of core activities and value creation outside of the EU.

---

[25] See n. 10.
[26] Academy for Artificial Intelligence in the German AI Association, Large AI Models for Germany [Große KI-Modelle für Deutschland], 2022, 56.
[27] Damien Geradin, Theano Karanikioti and Dimitrios Katsifis, 'GDPR Myopia: how a well-intended regulation ended up favouring large online platforms' (2021) 17 European Competition Journal 47.

From this perspective, the AI Act emerges as a crucial competition policy instrument in disguise: significant compliance costs, lack of safe harbours, and legal uncertainty may engender an increasing oligopolisation of an already concentrated market and can both endanger Europe's technological independence and ultimately cost consumers dearly. All of this should be considered when balancing the risks and opportunities of AI in regulatory terms.

## V. Critique and policy proposals

This economic and international background provides a foil for critically assessing the current proposals for the AI Act, and for suggesting policy amendments. Overall, it should be stressed that the AI Act is, in my view, a step in the right direction, especially with regard to transparency of AI use, documentation, continuous risk management, and data quality. However, there is still considerable need for improvement, particularly with respect to: the definition of AI; the concrete classification as a high-risk system; biometrics; the regulation of foundation models and generative AI; the AI value chain; the fundamental rights impact assessment; notice and action mechanisms; codes of conduct; and technical standards (below, 1.-10.).

### 1. The Definition of AI

For any piece of legislation specifically targeting AI, the concept of AI itself is of obvious importance, delineating the scope of application. In the European context, this holds not only for the AI Act but also the Product Liability Directive (PLD) and the currently parked AI Liability Directive (AILD). Both would introduce significant and entirely novel evidence disclosure mechanisms and reversals of the burden of proof in the European (PLD) and national (AILD) product liability and tort frameworks.[29] With respect to the concept of AI, both frameworks refer to the AI Act.

In its original proposal, the EU Commission sought to delineate AI systems by reference to a list of specific computer science techniques in the former Annex I. Beyond the currently mostly used machine learning frameworks, this also included the older knowledge-based approaches[30] as well as "[s]tatistical approaches, Bayesian estimation, search and optimization methods". The inclusion of these latter approaches, particularly, may be criticized as overly broad as they form part and parcel of many established software solutions that have little in common with advanced AI–and the risks raised by it. While one may consider extending the scope of the AI Act to opaque or highly complex non-AI software,[31] civil rights organizations, however, have continued to push for a broad definition of AI to "future proof" the act and close any loopholes.

---

[28] Cf. Victor Fleischer, 'Regulatory arbitrage' (2010) 89 Tex L Rev 227; Greg Buchak and others, 'Fintech, regulatory arbitrage, and the rise of shadow banks' (2018) 130 Journal of Financial Economics 453.

[29] See, e.g., Hacker, 'The European AI Liability Directives - Critique of a Half-Hearted Approach and Lessons for the Future'; Gerhard Wagner, 'Liability Rules for the Digital Age - Aiming for the Brussels Effect' (2022) European Journal of Tort Law 191.

[30] See, e.g., Ram D Sriram, *Intelligent systems for engineering: a knowledge-based approach* (Springer 2012).

The Council and Parliament have sharpened the definition, particularly by homing in on the concept of "autonomy". However, that concept is now, in turn, insufficiently defined. First, the definition states that an AI system may be endowed "with various levels of autonomy" (Art. 3(1) AI Act EP Version), without specifying which degree is sufficient. Second, autonomy is supposed to mean that AI systems "have at least some degree of independence of actions from human controls and of capabilities to operate without human intervention".[32] This could potentially include many technologies of little relevance to the Act and the risks it addresses, including smart meters, scheduling tools, rule-based systems and almost any advanced software. An electric toothbrush can operate without human intervention, even though it clearly does not feature any AI (yet).

Hence, to tighten the focus of the AI Act, models should be required to have some ability to learn or adapt to new environments. Independence from human control and interaction is not sufficient for autonomy. Rather, it defines machines that are *automated* but not autonomous.[33] To specifically focus on AI systems, and the specific risks they entail, it makes sense to add an "ability to learn and/or adapt autonomously to new environments" as a prerequisite. Otherwise, the AI Act does become a Software Act–which would have to be structured and written in a different way.

## 2. Classification as a high-risk system

Following the fundamentally use-case-centred architecture of the AI Act, high-risk systems are defined by reference to certain fields of application, listed exhaustively in Annexes II and III. These include applications in administration, employment, medicine, or biometrics, for example. Importantly, only systems classified as high-risk need to comply with the bulk of legal obligations contained in the AI Act.

However, both the Council and the Parliament have made proposals to differentiate further. A so-called *extra layer* was inserted in Art. 6 AI Act, according to which merely accessory applications in high-risk areas are excluded. The Parliament has suggested to apply the high-risk rules only if the concrete application indeed poses a significant risk (measured by probability and extent of harm). The trilogue also seems to be moving in this direction.

In my view, this makes sense for two reasons. First, completely different AI risk profiles do indeed coexist within high-risk areas: It does not seem convincing, especially with a risk-based approach, to subject an AI for medical operations to the same requirements as an AI that merely manages the doctor's appointments. Second, the concept of AI and thus the scope of the regulation is, as seen, quite broad. Without the extra layer, that would be a risk that even simpler software used for an administrative activity in high-risk areas might fall under the AI regulation and be difficult to use, despite obvious benefits and low risk.

---
[31] Hacker, 'The European AI Liability Directives - Critique of a Half-Hearted Approach and Lessons for the Future', 9.
[32] Recital 6 AI Act EP Version.
[33] See Stuart J. Russell and Peter Norvig, Artificial Intelligence: A Modern Approach (3rd Global ed. edn, Pearson Education, Inc. 2016) 39; Catherine Tessier, 'Robots autonomy: Some technical issues' in W.F. Lawless and others (eds), *Autonomy and Artificial Intelligence: A Threat or Savior?* (2017), 180.

### 3. Biometrics

The subject of live remote biometric identification, such as facial recognition technology in public spaces, sits at the intersection of prohibited and high risks. It has proven to be a highly contentious issue within the framework of the AI Act, almost derailing a hard-fought compromise during deliberations in the European Parliament, giving rise to intense debates and lobbying efforts. Two main camps emerged: one advocating for the complete elimination of real-time remote biometric identification in public spaces, citing concerns over civil liberties, data privacy, and the potential for mass surveillance.[34] The opposing camp argued for narrow exceptions to the rule, specifically in instances like crime prevention, prosecution, and matters of national security.[35]

In my view, the European Commission's original proposal, which allowed for narrow exceptions in cases of overriding public interest, strikes a more balanced approach. While concerns about the abuse of live remote biometric identification technologies are legitimate, there are scenarios where their use could be both legitimate and beneficial. For example, in cases of missing children or imminent terrorist threats, the technology could prove invaluable for rapid identification and response. This would necessitate stringent oversight and regulation to ensure it is not misused or expanded beyond these exceptional circumstances (function creep[36]). However, an outright ban would potentially put a significant number of innocent persons in harm's way in these critical situations. Therefore, narrow exceptions should remain possible under the Act, guided by clear protocols and checks to safeguard against abuse.

Perhaps equally important, the fierce discussion on biometric surveillance tends to obscure a wider and largely underappreciated problem: biometrics sits at the heart of augmented reality, a technology that is already significantly impacting the way consumers search and shop online, and may drive interaction with technology in the future.[37] Note that augmented reality is different from a purely virtual metaverse and consists in blending virtual and real content.[38] Prominent, and societally useful, applications include not only cultural enactments and educational purposes but also, for example, applications helping online customers choose the right size for clothing and accessories.[39] This, in turn, helps lower product returns, which is crucial both from an economic and sustainability perspective. However, determining the right customer size quite obviously depends on exact body measurements. In my view, there is a clear societal interest in keeping such functionalities open if the respective user has validly consented.

---

[34] See, e.g., Irena Barkane, 'Questioning the EU proposal for an Artificial Intelligence Act: The need for prohibitions and a stricter approach to biometric surveillance 1' (2022) 27 Information Polity 147.
[35] See also Margaret Hu, 'Biometrics and an AI Bill of Rights' (2022) 60 Duq L Rev 283.
[36] Cf. Bert-Jaap Koops, 'The concept of function creep' (2021) 13 Law, Innovation and Technology 29.
[37] Scott G Dacko, 'Enabling smart retail settings via mobile augmented reality shopping apps' (2017) 124 Technological Forecasting and Social Change 243; Yi Jiang, Xueqin Wang and Kum Fai Yuen, 'Augmented reality shopping application usage: The influence of attitude, value, and characteristics of innovation' (2021) 63 Journal of Retailing and Consumer Services 102720.
[38] Yuntao Wang et al. "A Survey on Metaverse: Fundamentals, Security, and Privacy." *arXiv preprint arXiv:2203.02662* (2022), 1-5; Will Greenwald, *Augmented Reality (AR) vs. Virtual Reality (VR): What's the Difference?*, PC MAG. (Mar. 31, 2021), https://www.pcmag.com/news/augmented-reality-ar-vs-virtual-reality-vr-whats-the-difference.
[39] See also Graeme McLean and Alan Wilson, 'Shopping in the digital world: Examining customer engagement through augmented reality mobile applications' (2019) 101 Computers in Human Behavior 210.

Hence, the regulation of biometric applications of AI should not throw the baby with the bathwater. Overly strict limitation of using AI-based innovation in this area could curtail the use of consensual and ephemeral measurements in public spaces like fashion boutiques. While the concept of biometric data, taken from Art. 4(14) GDPR, provides some necessary focus by relying on identification purposes,[40] regulators should be aware of the fact that biometrics, even in the sense of "real-time remote biometric identification", may have a legitimate economical use cases beyond the prevention or prosecution of severe crime.

### 4. The regulation of foundation models

The importance of biometrics is only surpassed by the other regulatory core issue: No other topic has sparked so much interest and controversy concerning the AI Act as the possible regulation of *foundation models* (FMs), such as ChatGPT, Claude (Anthropic), PaLM and Bard (Google) or LLaMA (Meta).[41] FMs are particularly potent models that have been trained on large amounts of data and form the basis for a wide range of downstream applications.[42] The Council de facto wanted to unequivocally designate such models as high-risk applications–and would thereby have turned the application-based architecture of the Act on its head.

In my view, three levels have to be distinguished in this context:[43] the regulation of the foundation models themselves; their applications; and the actors in the value chain. In principle, this architecture is now implemented in the AI Act EP version. While Art. 28b formulates a set of requirements at the model level, Art. 9 et seqq. Apply to specific high-risk applications; Art. 28 in turn regulates the value chain (see below, 6.).

In particular, it makes sense to set certain minimum standards at the model level, which responsible companies (should) fulfil, anyway. This concerns transparency with regard to the metadata of training data; due diligence in the selection of training data (concerning bias and representativeness); but also robust IT security and measures against the abuse of FMs for cyber and hybrid attacks (see below, VI.3.).[44] The latter, particularly, has become of the outsize importance with the fraud geopolitical environment the EU is currently navigating.

Art. 28b(2)(a) and (f) AI Act EP version, however, also require extensive risk assessments and risk management at the foundation model level. As we have argued elsewhere in detail, this is inefficient, unnecessary, and burdens SMEs in particular.[45] For example, an FM like GPT-4 (Open AI) or Luminous (Aleph Alpha, based in Heidelberg, Germany) could easily have 10,000 different use cases in high-risk scenarios. Requiring comprehensive risk management would force the FM developer to assess, mitigate and manage all these risks–even though only a fraction may come into play in real applications. Therefore, Article 28b(2)(a) should be reduced to a rough summary of possible risks in order to limit the actual risk management to specific high-risk cases, as Art. 9 AI Act provides for, anyway.

---

[40] See, e.g., Els J Kindt, 'Having yes, using no? About the new legal regime for biometric data' (2018) 34 Computer Law & Security Review 523.
[41] See, e.g., Lilian Edwards, 'Regulating AI in Europe: four problems and four solutions' (2022); Helberger and Diakopoulos, 'ChatGPT and the AI Act'; Hacker, Engel and Mauer, 'Regulating ChatGPT and other Large Generative AI Models'.
[42] See Art. 3 para. 1c AI Act EP version; based on Rishi Bommasani et al, 'On the opportunities and risks of foundation models' (2021) arXiv preprint arXiv:210807258, 2 ff.
[43] Hacker/Engel/Mauer, FAccT '23, 1112, 1114 et seq.
[44] See, e.g., the warning by the Dutch Cyber Security Center, AI: Cruciaal moment in de geschiedenis of een hype?, June 6, 2023, https://www.ncsc.nl/actueel/weblog/weblog/2023/ai-cruciaal-moment-in-de-geschiedenis-of-een-hype.

The scope and depth of risk management at the model level could, and should, ultimately depend on a range of factors: the capabilities of the model; the number of users; and the size (i.e., turnover and number of employees) of the developing company or its parent company. As Kai Zenner has rightly argued in a recent proposal, such a focus on "systemic foundation models" could transfer valuable lessons from the DSA regulatory architecture (special rules for "VLOPs") and exonerate SMEs.[46]

Nonetheless, in my view, non-systemic foundation models cannot be left entirely off the hook. While greater efforts can be expected from larger companies, highly capable and risky models can also be developed by small start-ups. In a risk-based approach, being small cannot be a *carte blanche* for putting unsafe products onto the market. Product safety regulation is quite uncompromising and clear in this respect, and for good reasons. Hence, even SMEs must fulfil minimum standards, but the compliance efforts that can reasonably be expected should take systemic elements, such as capability, users, and company size, into account.

### 5. The regulation of generative AI

In view of its considerable economic and social impact, the regulation of generative AI is another policy hotspot. Generative models are not identical with foundation models, even if the relevant foundation models that exist today are also generative. The distinguishing characteristic of the former is communicative content as output, such as text, images, videos or music.[47] Examples besides ChatGPT are LLaMA, DALLE, Midjourney or Stable Diffusion.

Art. 28b para. 4 AI Act EP Version lists three quite heterogeneous requirements. First, the use of generative AI models must be disclosed to users interacting with them (lit. a in conjunction with Art. 52), which makes sense to prevent or at least fine unnoticed impersonation. Arguably, however, persistent identifiers, such as invisible watermarks[48] or cryptographic identifiers, would be even more important so that anyone exposed to content, for example in social networks, can check whether it has been generated by AI. For example, China has recently launched a disinformation campaign concerning the Maui wildfires, claiming that the US tested whether bomb on its own population, backed up by AI generated pictures.[49] Second, the creation of illegal content must be prevented according to the AI Act's generative AI rules (lit. b). This points in the right direction, but does not go far enough (see below, 8.).

---

[45] Hacker/Engel/Mauer, FAccT '23, 1112, 1114 et seq.
[46] Kai Zenner, 'A law for foundation models: the EU AI Act can improve regulation for fairer competition' OECD AI Policy Observatory Blog (July 20, 2023) <https://oecd.ai/en/wonk/foundation-models-eu-ai-act-fairer-competition>.
[47] Art. 28b para. 4 AI Act EP version.
[48] Alexei Grinbaum and Laurynas Adomaitis, 'The Ethical Need for Watermarks in Machine-Generated Language' (2022) arXiv preprint arXiv:220903118; John Kirchenbauer and others, 'A Watermark for Large Language Models' (2023) arXiv preprint arXiv:230110226; Miranda Christ, Sam Gunn and Or Zamir, 'Undetectable Watermarks for Language Models' (2023) arXiv preprint arXiv:230609194.

Finally, a sufficiently detailed summary is envisaged on the use of copyright-protected training data (lit. c). The obligation to disclose copyrighted training data is a significant burden, especially because copyright law is not uniform across the EU and only partially harmonized via a contested line of CJEU jurisprudence.[50] The question of what counts as copyrighted material is often contentious, especially in the case of two-dimensional images.[51] Carrying out appropriate due diligence will hardly be feasible for developers who process large amounts of data (up to several billion training data points). In this respect, it would be better to create a moderated access for creators, through which they can selectively check whether training data contain material to which they have rights. However, such access should not lead to the disclosure of the entire training data material, as this would constitute an open invitation for competitors and strategic opponents to simply copy-paste highly valuable training data for free (*free riding*).

## 6. The AI value chain

AI is not always developed by a single company and then put into use. Particularly in settings involving foundation models, these pre-trained systems are often developed by one company (for example, OpenAI, Google or Aleph Alpha) and then fine-tuned by other entities for specific tasks, before being deployed at a specific end user.[52]

Ideally, the actor who can exert the most effective or least costly influence should be the accuracy of regulation. For this purpose, Art. 28(1) AI Act EP Version now provides for a transfer of obligations to those actors who substantially modify the AI system. Going beyond this role, I would submit that, in the case of changes agreed with the provider, the latter should also continue to be jointly and severally liable (cf. Art. 26 GDPR).

Another problem is that often none of the parties involved has all the information needed to fulfil the AI Act obligations: The developers (e.g. OpenAI) do not know how the model will be applied in situ; intermediaries and users, in turn, ignore the details of the training data. Therefore, access and information rights are necessary to ensure effective compliance. Art. 28(2) AI Act EP Version now also provides for this. Simultaneously, however, trade secrets and intellectual property must be protected in order to avoid competitive disadvantages for providers, especially with regard to competitors eager to free ride on valuable know-how, both from within the EU and from strategic rivals. There are various legal instruments for this, such as confidentiality agreements and non-competition clauses; protective orders by courts (cf. Article 3(4) of the draft AI Liability Directive AILD); or independent experts (Special Master, cf. US F.R.C.P. Rule 53(a)).[53] Art. 28(2b) AI Act EP version is still too vague here and should be amended.

---

[49] David E. Sanger and Steven Lee Myers, 'China Sows Disinformation About Hawaii Fires Using New Techniques' New York Times (September 11, 2023) <https://www.nytimes.com/2023/09/11/us/politics/china-disinformation-ai.html>.

[50] See, e.g., Raquel Xalabarder, 'The role of the CJEU in harmonizing EU copyright law' (2016) 47 IIC-International Review of Intellectual Property and Competition Law 635; Ana Ramalho, 'The Competence and Rationale of EU Copyright Harmonization" in Eleonora Rosati (ed), *The Routledge Handbook of EU Copyright Law* (2021), 3.

[51] Thomas Margoni, 'Digitising the public domain: non original photographs in comparative EU copyright law' in John Gilchrist and Brian Fitzgerald (eds), *Copyright, Property and the Social Contract* (Springer 2018), 157.

[52] See also Matt Bornstein, Guido Appenzeller and Martin Casado, 'Who Owns the Generative AI Platform?' Andreessen Horowitz Blog (January 19, 2023) <https://a16z.com/2023/01/19/who-owns-the-generative-ai-platform/> accessed February 6, 2023.

### 7. Fundamental rights impact assessment

The fundamental rights impact assessment newly introduced in Art. 29a AI Act EP Version should be reconsidered. While fundamental rights are the backbone of the rule of law, it is dogmatically unclear if and how companies can directly violate fundamental rights (horizontal direct effect).[54] Above all, however, Article 9 provides for a general risk assessment anyway, which also includes fundamental rights aspects. Finally, Article 35 of the GDPR requires a data protection impact assessment for personal data in critical scenarios, anyway. The added value of Art. 29a beyond the generation of standard boilerplate is therefore uncertain, unless re-conceptualized in a much more ambitious way as a tool for stakeholder participation.[55]

### 8. Notice and action mechanism

More importantly, the DSA is the EU's flagship regulation aiming to stem the tide of hate speech and fake news–but it does not directly apply to generative AI platforms.[56] While posting AI-generated content on covered intermediaries, such as Twitter/X or Facebook, does trigger DSA duties for these platforms, experience shows that the harm created can hardly be undone. Hence, AI-based illegal content should be tackled at its source, i.e., generative AI itself. Expanding DSA duties to AI providers would create a clear and manageable framework, for example by introducing a mandatory notice and action mechanism (Article 16 DSA), and endowing decentralized red teaming with the priority status of trusted flaggers (Article 22 DSA).[57] These measures would decentralise control over AI outcomes, draw on the monitoring resources of civil society, and ensure a safer AI ecosystem.

### 9. Potentially binding codes of conduct

Regulators and civil society alone, however, will not tame the risks of advanced AI systems. Rather, industry collaboration and implementation will be crucial. Hence, the potential of regulated self-regulation should be exploited.[58] In the GDPR, Art. 40, for example, allows associations to not only develop codes of conduct but also to have them approved by supervisory authorities.[59] This possibility should also be made available for codes of conduct in the area of AI; Art. 69 AI Act should be updated supplemented in this regard.

---

[53] On this in greater detail Hacker/Engel/Mauer, FAccT '23, 1112, 1117.
[54] See, on this long-standing debate, Eleni Frantziou, *The Horizontal Effect of Fundamental Rights in the European Union: A Constitutional Analysis* (Oxford University Press 2019); Nuria Bermejo, 'Fundamental Rights and Horizontal Direct Effect under the Charter' in Cristina Izquierdo-Sans, Carmen Martínez-Capdevila and Magdalena Nogueira-Guastavino (eds), *Fundamental Rights Challenges: Horizontal Effectiveness, Rule of Law and Margin of National Appreciation* (Springer 2021), 51.
[55] In this sense the open letter drafted by Malgieri et al.
[56] On this in more detail: Hacker/Engel/Mauer, FAccT '23, 1112, 1117 f.
[57] Ibid.
[58] Julia Black, 'Decentring Regulation: Understanding the Role of Regulation and Self-Regulation in a 'Post-Regulatory' World' (2001) 54 Current Legal Problems 103; Wolfgang Schulz and Thorsten Held, *Regulated self-regulation as a form of modern government* (Study Commissioned by the German Ferderal Commissioner for Cultural and Media Affairs 2001).

More specifically, to be eligible, such proposals should be endorsed by at least one trade association acting in the specific sector, or by a minimum number of AI companies operating in a specific area (e.g., 10). Regulatory approval would endow general validity and thus offer developers and providers a *safe harbour*. This seems essential, especially in the context of the challenging liability scenarios described above. Developers and deployers would still be able to operate outside of such safe harbours, but at their own risk.

### 10. Setting technical standards

Another key feature for reducing legal uncertainty and providing safe harbours are technical standards and common specifications mentioned abstractly in Art. 40 and 41 AI Act.[60] Crucially, these standards should be used case specific and provide clear guidelines (e.g.: "the F1 score for use case X must be at least Y") to effectively link abstract obligations and practical implementation. Corresponding ISO[61] and CEN-CENELEC standards[62] are in progress. Much will depend on the exact scope and shape of these standards. Ideally, they should be composed of both procedural and substantive components. Such standards would offer further safe harbours, for the benefit of small and large AI developers, and compliance in general.

### VI. Future regulatory challenges in AI

Many risks can, and will, be addressed in the AI Act. However, this is not the end of the story: AI itself, and its regulation, will probably be with us for quite a while. Therefore, it may be permissible to survey some additional risks that have not been sufficiently addressed in the AI Act, but that are increasingly shaping the impact of AI on society.

Advanced AI systems are marked by three characteristics: they are often generative, producing communicative output; they are trained on large data at scale, necessitating vast amounts of compute, specialized chips, and resources; and they are increasingly capable to effectively execute a variety of hard goals and complex instructions. From these criteria, three critical "AI externalities" can be derived: misinformation; environmental costs; and hybrid threats leveraging highly capable AI systems. In my view, these three risks will take on particular urgency and should be at the centre of any current or future AI policy and research agenda. All three constitute different types of AI externalities that technology, for all its benefits, unloads on society.

---

[59] See for example Hacker, Datenprivatrecht, 2020, 307 ff.
[60] See, e.g., Hans-W. Micklitz, *The Role of Standards in Future EU Digital Policy Legislation* (ANEC/BEUC 2023).
[61] ISO (International Organization for Standardization), Standards by ISO/IEC JTC 1/SC 42
 Artificial Intelligence, https://www.iso.org/committee/6794475/x/catalogue/p/0/u/1/w/0/d/0.
[62] CEN-CENELEC (European Committee for Standardisation and European Committee for Electrotechnical Standardisation), Artificial Intelligence, https://www.cencenelec.eu/areas-of-work/cen-cenelec-topics/artificial-intelligence/.

1. **Toxic content**

The automated production of misinformation has been recognized as a key challenge in generative AI systems. Fake news and hate speech are not merely theoretical by-products of advanced AI systems. As a recent study have shown, ChatGPT et al. can be used for the mass generation of professionally crafted hate speech, including computer code for the most efficient social media distribution.[63] Despite efforts by responsible AI developing companies to rein in such rampages, moderation policies are anything but watertight, and jailbreak prompts abound.[64] As a recent experiment has shown, a leading European generative AI model ends the prompt "Muslims are..." with "... the enemies of humanity," and other derogatory terms.[65] New avenues beyond DSA-style content moderation need to be explored here, starting with mandatory training and testing, and a rigorous rethinking of the balancing of freedom of speech on the one hand and the protection of critical societal infrastructures such as elections and climate adaptation strategies on the other hand.[66]

2. **Environmental costs**

Increasingly, computer scientists are sounding the alarm on the massive energy and water consumption of advanced AI models, both in training and deployment.[67] However, the legal reflection on these considerable environmental costs is still in its infancy.[68] According to current estimates, information and communication technology (ICT) contributes up to 3.9 % to global greenhouse gas emissions at[69] - compared to about 2.5 % for global air travel.[70] The carbon footprint of machine learning in particular has increased significantly in recent years.[71] Large models used for generative AI are bound to exacerbate this trend.[72]

---

[63] Beuth, How ChatGPT can be hacked with words. Der Spiegel (Jan. 12, 2023), https://www.spiegel.de/netzwelt/web/chatgpt-wie-sich-die-kuenstliche-intelligenz-mit-worten-hacken-laesst-a-2a3dd1b4-7405-40e0-8ba0-172915f38e57.
[64] Yi Liu and others, 'Jailbreaking chatgpt via prompt engineering: An empirical study' (2023) arXiv preprint arXiv:230513860.
[65] See von Lindern, Braucht die deutsche Vorzeige-KI mehr Erziehung?, ZEIT Online (Sept. 11, 2023), https://www.zeit.de/digital/2023-09/aleph-alpha-luminous-jonas-andrulis-generative-ki-rassismus, referring to Aleph Alpha's Luminous model.
[66] Amelie Berz, Andreas Engel and Philipp Hacker, 'Generative KI, Datenschutz, Hassrede und Desinformation – Zur Regulierung von KI-Meinungen' (2023) Zeitschrift für Urheber- und Medienrecht 586, 593 et seq.
[67] Charlotte Freitag and others, 'The real climate and transformative impact of ICT: A critique of estimates, trends, and regulations' (2021) 2 Patterns 100340; Bran Knowles and others, 'Our house is on fire: The climate emergency and computing's responsibility' (2022) 65 Communications of the ACM 38.
[68] But see Ugo Pagallo, Jacopo Ciani Sciolla and Massimo Durante, 'The environmental challenges of AI in EU law: lessons learned from the Artificial Intelligence Act (AIA) with its drawbacks ' (2022) 16 Transforming Government: People, Process and Policy 359; Philipp Hacker, 'Sustainable AI Regulation' (September, 2023) Working Paper, presented at the Privacy Law Scholars Conference 2023, https://arxivorg/abs/230600292.
[69] Freitag et al., 'The real climate and transformative impact of ICT: A critique of estimates, trends, and regulations'; OECD, Measuring the Environmental Impacts of AI Compute and Applications, 2022, 25-26.
[70] ACM Tech. Policy Council, *ACM TechBrief: Computing and Climate Change* (2021) 1.
[71] Ibid: possibly by a factor of up to 300,000.
[72] OECD, Measuring the EnvironmentalImpacts of AI Compute and Applications, 2022, 5; Li and others, 'Making AI Less "Thirsty": Uncovering and Addressing the Secret Water Footprint of AI Models', Working Paper, 2023, 2.

Transparency alone won't solve this challenge. The European Parliament has included several provisions in its version, ranging from general principles and sustainability transparency to environmental risk assessments. Beyond this, conceivable options include sustainability by design, for example through sustainability impact assessments as part of AI development; resource-conserving specifications for AI training; and finally, the inclusion of AI processes and AI infrastructures in EU emissions trading.[73]

### 3. Hybrid threats

Finally, addressing these two risks is complicated by the emergence of increasingly powerful open-source AI models that can be downloaded and used for free. Specifically, large language models (LLMs) are advancing at an accelerated pace and are commonly viewed as a foundational technology for future autonomous AI agents.[74] Open-source software libraries make these agents easily accessible even to those without specialized knowledge.[75] With AI systems poised to augment their functionalities substantially in the near future, they could become powerful tools in the hands of potentially malicious actors.[76] The Dutch Cyber Security Centre has issued a resounding warning concerning the weaponization of LLMs, noting that they may be harnessed "for setting up complex cyberattacks, writing exploit code for unknown security issues or converting published vulnerabilities to potent exploit software, at least. Both in the planning and execution phases, everything can be automated."[77] Beyond this, advanced systems may be used for hybrid threats, i.e., simultaneous attacks and physical and cyberspace, for example, on critical infrastructure.[78]

Therefore, urgent measures are needed to establish protocols for testing and controlled access to high-performance AI systems; such regulations may eventually necessitate restrictions on their open-source availability, including an open-source ban for particularly capable systems. Unfortunately, the path from ChatGPT to "ThreatGPT" and is straightforward and rather short.[79] Therefore, rules on mandatory threat-testing and moderated access of frontier AI systems should urgently be included in any AI regulation–in the EU and beyond.

## VII.   Conclusion

The debate on AI regulation in the EU has once again exposed vastly different regulatory philosophies, ranging from sectoral and self-regulation approaches in the UK to encompassing horizontal command-and-control legislation paired with ramped-up liability in the EU. In my view, a mixed strategy will be necessary, drawing on all elements is necessary, and particularly providing agility and safe harbours to facilitate compliance. Two other regions of the world contemplating AI regulation, this will serve as a reminder that technical and economic questions are inherently interwoven with regulatory strategies and philosophies.

---

[73] In detail Philipp Hacker, 'Sustainable AI Regulation' (September, 2023) Working Paper, presented at the Privacy Law Scholars Conference 2023, https://arxivorg/abs/230600292, 21 ff.
[74] Zhiheng Xi and others, 'The Rise and Potential of Large Language Model Based Agents: A Survey' (2023) arXiv preprint arXiv:230907864, 8-9.
[75] Wangchunshu Zhou and others, 'Agents: An Open-source Framework for Autonomous Language Agents' (2023) arXiv preprint arXiv:230907870.
[76] Harry Law and Sébastien Krier, 'Open-source provisions for large models in the AI Act' (2023) 4 Cambridge Journal of Science & Policy 1.
[77] Dutch Cyber Security Center, AI: Cruciaal moment in de geschiedenis of een hype?, June 6, 2023, https://www.ncsc.nl/actueel/weblog/weblog/2023/ai-cruciaal-moment-in-de-geschiedenis-of-een-hype [automated translation by Firefox].
[78] Carlos Pedro Gonçalves, 'Cyberspace and Artificial Intelligence: The New Face of Cyber-Enhanced Hybrid Threats', *Cyberspace* (IntechOpen 2019), ch 5.
[79] Maanak Gupta and others, 'From ChatGPT to ThreatGPT: Impact of generative AI in cybersecurity and privacy' (2023) 11 IEEE Access 80218.

The AI Act itself is a first attempt by the EU to address the multiple challenges of AI technology. Overall, I believe that the draft law goes in the right direction. However, shortcomings still threaten to hamper the development of generative AI in the EU and beyond. Ultimately, for example, risk management needs to be clearly aligned with a system's use case.

Finally, the chapter anticipates future regulatory challenges in AI, ranging from the management of toxic content and environmental sustainability to the issue of hybrid threats. It argues for urgent measures to establish protocols for controlled access to high-performance AI systems, particularly in the context of open-source models. The chapter concludes that while the EU's AI Act is a monumental step, it requires further refinement and international cooperation to effectively manage the complex landscape of AI technologies.